\documentclass[twoside]{article}
\usepackage{fleqn,espcrc2,amsmath}

\title{\begin{center}Geometrical Approach to the Gauge Field Mass Problem.\\
Possible Reasons for which the Higgs Bosons are
Unobservable.\end{center}}

\author{Yu.P. Peresun'ko \address{NSC KIPT, 61108, Kharkov, Ukraine.\\
 E-mail: peresunko@kipt.kharkov.ua}}

\begin{document}
\setcounter{page}{391}
\begin{abstract}
   The (4+d)-dimensional Einstein--Hilbert
   gravity action is considered in the Kaluza - Klein  approach.
   The extra-dimensional manifold $V_d$
   is a Riemannian space with the $d$-parametric group of isometries $G_d$
which
    acts on $V_d$ by the left shifts and with an arbitrary
    non-degenerate
   left-invariant metric $\dot{g}_{ab}$. The gauge fields
$\widehat{A}_\mu(x)$
    are introduced as the affine connection coefficients of the fiber bundle
    with $V_d$ being the fiber. The effective Lagrangian
    $L_{\text{eff}}\{\widehat{A}_\mu(x),\dot{g}_{ab}\} $ is obtained
    as an invariant integral of the curvature scalar of the structure
considered.
    The conditions on $\dot{g}_{ab}$ are formulated under which
    $L_{\text{eff}}\{\widehat{A}_\mu(x),\dot{g}_{ab}\}$ contains in addition
to the
    square of the gauge field strength tensor also the quadratic form of
    $\widehat{A}_\mu(x)$ and additional fields with pure gauge degrees of
    freedom. The eigenvalues of the quadratic form
    are calculated for the case of the gauge group $SO(3)$ and it is shown
that they
    are not equal to zero in the case when $\dot{g}_{ab}$ is not
    proportional to the unit matrix.
     \vspace{1pc}
\end{abstract}

\maketitle

As it is known, the most consistent approach to the theoretical
description of the gauge fields is using Kaluza - Klein type
theories of (4+d)-dimensional Einstein gravity
(\cite{1}-\cite{3}). In this approach the properties of the gauge
fields are the consequence of the geometrical and topological
structure of the extra-dimensional manifold. The effective
Lagrangian of this theory is obtained upon integration over the
extra-dimensional manifold and, depending on the type of this
space, can contain gauge fields, as well as various fermionic and
scalar fields.

 There is a well known difficulty in the description of massive gauge fields
$\widehat{A}_\mu (x)$ which consists in the following. The
quadratic form $\widehat{A}_\mu(x)\widehat{M}\widehat{A}_\mu(x)$
which describes mass terms in a 4-dimension Lagrangian is not
invariant under inhomogeneous transformations of the gauge fields
$\widehat{A}_\mu (x)\rightarrow \widehat{A}_\mu ^{\prime}(x)
     =S(x)\widehat{A}_\mu (x)S^{-1}(x)+S(x)\partial _\mu S^{-1}(x)$.
 As is well known, this problem is solved by the introduction
into the theory of additional scalar fields with an appropriate
transformation law and with such a self-interaction potential in
the ordinary space-time which cause the scalar field  to acquire a
non-zero vacuum expectation value (VEV). The interaction of the
gauge fields with this VEV produces mass terms for gauge fields.
(the Higgs effect). This mechanism of the VEV generation should be
necessarily accompanied by the quantum excitations over the VEV.
The absence of the experimental observations of such excitations
(Higgs bosons) impels one to think that it would be more natural
to introduce into the theory of an intrinsic analog of the VEV
object in the same manner as charges, lepton masses etc.  are
introduced. The origin of such objects should be regarded as the
subject for a future study of the theory. In this work I would
like to show that the metric of the extra-dimensional manifold in
the Kaluza-Klein approach may be used as such an object.

 We will shown  that in
multi-dimensional theories of the Kaluza-Klein type one can obtain
the Lagrangian which is manifestly gauge invariant in
multi-dimensional space-time and after the integration over the
additional manifold reduces to an in effective 4-dimensionl
Lagrangian which contains both the square of the strength tensor
of gauge fields and the quadratic form of this fields. Note that
a selection of parameters of the metric of $V_d$ is possible such
that corresponding additional fields are pure gauge degrees of
freedom and hence unobservable.

 Let us briefly remind the basic structure of Kaluza - Klein type theories.
 The effective 4-dimensional action is obtained from these theories after
the
 integration over the extra-dimensional manifold of the (4+d)-dimensional
 Einstein--Hilbert action:
\begin{equation}\label{eq1}
S=\frac 1{\kappa ^2}\int d^{(4+d)}X\sqrt{-G} R = \int
d^4xL_{\rm{eff}}(x)
\end{equation}
where $\kappa $is a (4+d)-dimensional gravitational constant,
$G, R$ are a metric  and the scalar curvature in
(4+d)-dimensional space. The coordinates $X^A$,
$A=0,1,2,3,...,3+d,$ are separated into the coordinates $x^{\mu
}$, $\mu =0,...,3$, of the ordinary space--time, plus the
coordinates
$y^{\alpha }$, $\alpha =1,2,...,d$, of a compact d-dimensional
manifold $V_d$.

We shall restrict our consideration to the Yang-Mills gauge fields
with a $r$ - parametric gauge group $G_r$, without taking into
account effects of the gravitational field in ordinary
space--time. In other words, we shall assume that
(4+d)-dimensional space has a structure of the fiber bundle with
the flat Minkowski space $M_4$ (with a metric
 $\gamma _{\mu \nu}=$ diag$(1,-1,-1,-1)$) being its base and the
 fiber $V_d$ being a Riemannian space with the $r$-parametric group of
 isometries $G_r$ which acts transitively on $V_d$ by the left shifts. This
means
 that the dimension of $V_d$ (i.e. $d$) is equal to $r$. Yang-Mills gauge
fields are
 introduced as the affine connection coefficients
\begin{equation}\label{eq2}
  \widehat{A}_\mu (x)=A_\mu ^a(x)E_a(y).
\end{equation}
Here $E_a (y)$, $a=1,...,r$ are generators of the left shifts on
$V_d$ which obey the Lie algebra of $G_r$:
\begin{equation}\label{eq3}
 \left[ E_a,E_b\right] =f_{ab}^cE_c,
 \end{equation}
where $f_{ab}^c$ are the structure constants of $G_r$, and $E_a$
have the form:
\begin{equation}\label{eq4} E_a(y)=\xi _a^\alpha (y)\partial /\partial
y^\alpha,
\end{equation}
 $\xi _a^\alpha (y)$ are corresponding Killing  vectors.
Hereinafter it is convenient to use the variables $y^\alpha $,
$A_{\mu}^a$, $g_{ab}$, $\gamma _{\mu \nu }$ and $E_a$ in the
dimensionless form. For this it is necessary to introduce a
constant $m^{-1}$ with the dimension of length.

Then in the covariant basis ${E_A}$, where
\begin{equation}\label{eq5}
\left\{
\begin{array}{lrl}
E_A=D_\mu \equiv \partial _\mu +m e A_\mu ^a E_a \;
 &A\!\!&=0,...,3 \\ E_A=m E_a\quad &A\!\!&=3+a \end{array} \right.
\end{equation}
the metric tensor $G_{A B }$ of the (4+d)-dimensional manifold
has the form:
\begin{equation}\label{eq6}
 G_{AB}=\left( \begin{array}{ll} g_{ab} & 0 \\ 0 &
\gamma _{\mu \nu } \end{array} \right)
\end{equation}
The commutation relations for the vector fields $E_A$ are
$[E_A,E_B]=C_{AB}^D E_D$, where
\begin{equation}\label{eq7}
\left\{
\begin{array}{l}
C_{AB}^D=eF_{\mu \nu }^d\mbox{ \hspace{0.1cm}when}\;A=\mu ,B=\nu
,D=d;
\\
C_{AB}^D=mf_{ab}^d \mbox{\hspace{0.1cm}when}\; A=a,B=b,D=d; \\
C_{AB}^D=0\mbox{ \hspace{0.15cm} \quad  otherwise.}
\end{array}
\right.
\end{equation}
 Here \\
 $F_{\mu \nu }^a=\partial _\mu A_\nu ^a(x)-\partial _\nu A_\mu
^a(x)+e m f_{bc}^a A_\mu ^b(x) A_\nu ^c(x)$ is the strength
tensor of the gauge field, $e m$ is the gauge field coupling
constant.

Using known formulae for the Christoffel symbols \cite{4}
\begin{eqnarray}\label{eq8}
\Gamma _{AB}^C&=&\frac
12G^{CD}(E_AG_{BD}+E_BG_{AD}-E_DG_{AB}\nonumber\\
&-&C_{AD}^EG_{BE}-C_{BD}^EG_{AE})+\frac 12C_{AD}^C,
\end{eqnarray}
and for the Riemann tensor
\begin{eqnarray}\label{eq9}
R_{\ BCD}^A&=&C_{DC}^E\Gamma _{EB}^A-E_D\Gamma _{CB}^A+E_C\Gamma
_{DB}^A \nonumber\\ &&-\;\Gamma _{DE}^A\Gamma _{CB}^E+\Gamma
_{CE}^A\Gamma _{DB}^E,
\end{eqnarray}
one can obtain the following expression for the scalar curvature
\cite{4}:
\begin{eqnarray}\label{eq10}
R=&-&\frac {e^2}{4}g_{ab}\gamma ^{\mu \rho }\gamma ^{\nu \sigma
}F_{\mu \nu }^aF_{\rho \sigma }^b
\nonumber\\&&+\;R^{(b)}+R^{(f)}+R^{(M)}
\end{eqnarray}
  where $F_{\mu \nu }^a$ is defined in (\ref{eq7}),
\begin{equation}\label{eq11}
R^{(f)}=\frac {m^2}{2}(f_{am}^nf_{nb}^mg^{ab}+\frac
12f_{ad}^nf_{cb}^mg^{cd}g_{mn}g^{ab} )
\end{equation}
is the scalar curvature of the fiber.  $R^{(b)}=0$ in the case of
the flat base,
\begin{multline}\label{eq12}
R^{(M)}=\\ -\frac 12\gamma ^{\mu \nu }
  \Bigl\{g^{ab}(D_\mu( D_\nu g_{ab}) )
 +D_\mu ( g^{ab}(D_\nu g_{ab}) )  \\
 -\frac12g^{bd}g^{ac} \bigl[ ( D_\mu g_{ad}) ( D_\nu
g_{cb}) \\
 \qquad \qquad-( D_\mu
g_{ac}) ( D_\nu g_{bd})\bigr]\Bigr\}
\end{multline}

In order for when acting by the left shifts $E_a (y)$ on $V_r$
the group $G_r$ to be the isometry group the metric $g$ on $V_r$
must obey the Killing equation:
\begin{equation}\label{eq13}
\xi _a^\sigma \partial _\sigma \widetilde{g}_{\alpha \beta
}+\widetilde{g} _{\alpha \sigma }\partial _\beta \xi _a^\sigma
+\widetilde{g}_{\beta \sigma }\partial _\alpha \xi _a^\sigma =0
\end{equation}
where
$\widetilde{g}_{\alpha \beta}=g(\partial_\alpha,\partial_\beta )$
is the metric tensor in the coordinate basis, and
$\xi_a^\alpha$ are the Killing vectors of the left shifts $E_a$ defined in
(\ref{eq4}).

As it is known \cite{5}, the solution of (\ref{eq13}) in the basis
$E_a$ has the form
\begin{equation}\label{eq14}
g(E_a,E_b)=g_{ab}(y)=K_{aa_1}\dot{g}_{a_1a_2}\overline{K}_{b_1b},
\end{equation}
where $\dot{g}_{ab}=g(E_a^{(R)},E_b^{(R)})$ is the metric tensor
in the left invariant basis of the right shifts $E_a^{(R)}=\eta
_a^\alpha (y)\partial /\partial y_\alpha $, $a=1,2,...,r$; $
\left[ E_a^{(R)},E_b^{(R)}\right]=-f_{ab}^cE_c^{(R)}, \quad \left[
E_a,E_b^{(R)}\right] =0$.

$ \dot{g}_{ab}$ is an arbitrary symmetrical non-degenerate
$(r\times r)$ matrix with elements which do not depend on
$y_\alpha $. In eq. (\ref{eq14}) $K_{ab}(y)=\xi _a^\alpha
(y)\left( \eta _b^{-1}(y)\right) _\alpha $ is the matrix of the
$Ad$-representation of $G_r$, $\overline{K}_{ab}=K_{ba}$
is the transposed matrix, and it is always possible to put
det(K)=1.

The effective 4-dimensional Lagrangian has the form
\begin{equation}\label{eq15}
L_{\text{eff}}(x)=\int\nolimits_{V_r}d\mu (y)R
\end{equation}
where the left-invariant measure is\\
 $d\mu(y)=m^{-r}d^ry\sqrt{\widetilde{g}}=m^{-r}d^ry\sqrt{\dot{g}}/
\text{det}(\xi _a^\alpha )$.

$\widetilde{g}$ and $\dot{g}$  are the metric determinant in the
coordinate and left-invariant basis, respectively.

It should be mentioned here that operators of the group $G_r$ act
on $g_{ab}$ only by means of the generators $E_a$ acting on
$K_{ab}(y)$
\begin{equation}\label{eq16}
\left( E_sK_{ab}\right) =f_{sa}^rK_{rb}.
\end{equation}
Therefore $g_{ab}$ transforms under the action of the generators
$E_a$ as follows
\begin{equation}\label{eq17}
E_cg_{ab}=f_{ca}^rg_{rb}+f_{cb}^rg_{ra}
\end{equation}
and the inverse tensor $g^{ab},$ \quad $g^{ab}g_{ac}=\delta
^{a}_{c}$ transforms as
\begin{equation}\label{eq18}
E_cg^{ab}=-f_{cr}^ag^{rb}-f_{cr}^bg^{ra}.
\end{equation}
It is easy to see that the expression (\ref{eq10}) is manifestly
invariant under the gauge transformations
\begin{multline}\label{eq19}
\widehat{A}_\mu (x) \rightarrow\widehat{A}_\mu ^{\prime}(x)=\\
S_{(E)}(x)\widehat{A}_\mu (x)S_{(E)}^{-1}(x)+S_{(E)}(x)(\partial
_\mu S_{(E)}^{-1}(x)),
\\
 \dot{g}\rightarrow\dot{g}^{\prime}=S_{(E)}(x)\dot{g}S_{(E)}^{-1}(x).
 \phantom{AAAAAAAA}
\end{multline}
 where  $S_{(E)}(x)=\exp \left( \omega ^a(x)E_a\right)$ is a local gauge
 transformation.

Since the metric tensor
 $g_{ab}(y)=K_{aa_1}\dot{g}_{a_1b_1}\overline{K}_{b_1b}$ is
expressed in terms of the $Ad$-representation matrices $K_{ab}$,
for further consideration it is convenient to pass to the matrix
form of the generators $E_s$ of the group $G_r$
$E_s\rightarrow (f_s)_{ab}$, where $(f_s)_{ab}$ is a matrix form of
the structure constants $f^b_{sa}=(f_s)_{ab}$.

Then the expression $(D_\mu g_{ab})$ may be rewritten as follows
\begin{multline}
(D_\mu g_{ab})=\\
\partial _\mu \left(
K(y)\dot{g}\overline{K}(y)\right)_{ab}
+em(\widehat{A}_\mu)_{ac}(K(y)\dot{g}\overline{K}(y))_{cb}\\
-em(K(y)
\dot{g}\overline{K}(y))_{ac}(\widehat{A}_\mu )_{cb},
\end{multline}
where $(\widehat{A}_\mu)_{ab}=(A_\mu ^sf_s)_{ab}$.

After substituting this expression into (\ref{eq10}) -
(\ref{eq15}) and some algebra the effective Lagrangian may be
written in the following form
\begin{equation}\label{eq21}
L_{\text{eff}}\{A_\mu ^s,\dot{g}_{ab}\}=L^{(F)}+L^{(M)}+L^{(f)},
\end{equation}
where
\begin{multline}\label{eq22}
L^{(F)}=\\-\frac{e^2}{4\kappa ^2}\int\nolimits_{V_r}d\mu (y)\left(
K(y) \dot{g}\overline{K}(y)\right) _{ab}\gamma ^{\mu \rho }\gamma
^{\nu \sigma }F_{\mu \nu }^aF_{\rho \sigma }^b
\end{multline}
\begin{multline}\label{eq23}
L^{(M)}=-\frac 1{2\kappa ^2}\int\nolimits_{V_r}d\mu (y) \\
\Bigl\{(em)^2\bigl[Sp\{\widehat{A}_\mu\widehat{A}_\mu\}\phantom{AAAAAAAAAAAA
AAAAA}
\\ -Sp\{\widehat{A}_\mu K(y) \dot{g}\overline{K}(y)\widehat{A}_\mu
\overline{K}^{-1}(y)\dot{g }^{-1}K^{-1}(y)\}\bigr]\\
 +em\bigl[Sp\{K(y)(\partial _\mu \dot{g})\dot{g}^{-1}K^{-1}(y)
\widehat{A}_\mu \}\phantom{AAAAAAAAAAAAA}\\
\phantom{AAAAA}-Sp\{\widehat{A}_\mu\overline{K}^{-1}(y)\dot{g}
^{-1}(\partial _\mu \dot{g})\overline{K}(y)\}\bigr]\\
 +2Sp\{\dot{g}^{-1}(\partial
_\mu^2\dot{g})\}+\frac12[Sp\{\dot{g} ^{-1}(\partial _\mu
\dot{g})\}]^2\phantom{AAAAAAAA}\\ -\frac 32Sp\{
\dot{g}^{-1}(\partial _\mu \dot{g})\dot{g} ^{-1}(\partial _\mu
\dot{g})\}\Bigr\};
\end{multline}
\begin{equation}\label{eq24}
L^{(f)}=\frac 1{\kappa ^2}\int\nolimits_{V_r}d\mu (y)R^{(f)};
\end{equation}
Here $\widehat{A}_\mu $ denotes the matrix $(A_\mu
^s(x)f_s)_{ab}$; \thinspace
 $\dot{g}$, $\dot{g}^{-1}$, $K$, $\overline{K}$, $K^{-1}$,
 $\overline{K}^{-1}$
denote corresponding matrices $\dot{g}_{ab}$, ... ,
$(\overline{K}^{-1})_{ab}$,
 and matrix multiplication is defined in the standard way.

The gauge transformation (\ref{eq19}) now may be written in the
usual matrix form\\
{$\left(S(x)\right)_{ab}=\left(exp(\omega^s(x)f_s)\right)_{ab}$.}

Let us take into account that the transformation
\\
$K_{ab}(y)\rightarrow
K_{ab}^{\prime}(y)=(S(x)K(y)S^{-1}(x))_{ab}=K_{ab}(y^{\prime })$
is equivalent to a change of the variables $y_\alpha
\rightarrow y^\prime_ \alpha $ and that the measure $d\mu (y)$ is
invariant under such transformations.

Then it is easy to see that eqs. (\ref{eq21}) - (\ref{eq24}) for
the effective Lagrangian are gauge invariant
\begin{multline}\label{eq25}
\widehat{A}_\mu (x) \rightarrow\widehat{A}_\mu
^{\prime}(x)=\\
S(x)\widehat{A}_\mu (x)S^{-1}(x)+S(x)(\partial _\mu
S^{-1}(x)),  \\
 \dot{g}\rightarrow\dot{g}^{\prime}=S(x)\dot{g}S^{-1}(x).
 \phantom{AAAAAAAAAA}
\end{multline}

The integration of (\ref{eq21}) over
$\int\nolimits_{V_r}d\mu (y)$ in the general case of an arbitrary
group $G_r$ is a rather difficult prolbem, but the integration of
the first term of
$L_{\text{eff}}(x)$ in (\ref{eq21}) is trivial when one uses known
orthogonality relations for the irreducible unitary
representations $K_{ab}^\Lambda (y)$
\begin{multline}\label{eq26}
\int\nolimits_{V_r}d\mu (y)K_{a_1b_1}^{\Lambda _1}(y)\overline{K}
_{a_2b_2}^{\Lambda _2}(y)=\\
\frac{v_r}{\text{dim}(\Lambda
_1)}\delta _{\Lambda _1\Lambda _2}\delta _{a_1a_2}\delta
_{b_1b_2}.
\end{multline}
Here the index $\Lambda $ numbers different irreducible
representations of the group $G_r$, $v_r=\int\nolimits_{V_r}d\mu
(y)$ is the volume of the manifold $V_r$ and  dim$(\Lambda)$ is
the dimension of $K^{\Lambda }_{ab}$.

Then we obtain
\[
L^{(F)}=-\frac{v_r}{\kappa ^2}\frac{e^2}{4r}Sp(\dot{g})F_{\mu \rho
}^aF_{\nu \sigma }^a\gamma ^{\mu \nu }\gamma ^{\rho \sigma }.
\]

 One can see that in the case of arbitrary dependence of the matrix
 $\dot{g}_{ab}$ on $x_\mu $, in addition to the gauge fields $A^a _\mu (x) $
 the effective Lagrangian
$L_{\text{eff}}\{A_\mu ^s,\dot{g}_{ab}\}$
 also contains  Brans--Dicke type
 fields (because $Sp(\dot{g})$ depends on $x_\mu $) and a
 set of scalar fields, with a complicated
 self-interaction potential, which  interact with the gauge fields.
 In the general case these scalar fields belong to the
different representations of $G_r$. The conditions can be
formulated under which this set of fields form certain
representations of $G_r$, and the
 Lagrangian of a standard Higgs type can be obtained. (See, for example,
 \cite{6a}).

 But there is more interesting opportunity, the one to consider a left
invariant
 metric $\dot{g}_{ab}$ as an intrinsic, independent of $x_\mu $,
 characteristic of the gauge fields $A_\mu ^a(x)$. More precisely,
 let us suppose that $\dot{g}_{ab}$ has the form
 $\dot{g}_{ab}=(S_0(x)\dot{\bar{g}}S_0^{-1}(x))_{ab}$, where
 $\dot{\bar{g}}_{ab}$ is a symmetrical matrix  independent of $x_\mu $ and
 $S_0(x)$ is an arbitrary gauge transformation matrix.

  It should be noted that  this
condition on $\dot{g}_{ab}$ is a direct generalization  of the
Einstein general relativity principle to  high dimensions.
Really, the Einstein general relativity principle is the statement
that the 4-dimension metric tensor $\gamma ^{\mu \rho }(x) $ which
is associated to the arbitrary gravitational field, may be locally
transformed to the flat Minkowski space tensor
$\overline{\gamma}_{\mu \rho}=\text{diag}(1,-1,-1,-1) $ by the
transformations of the gauge group. (In the case of gravity this
is the group of general covariant transformations.)

The generalization of this principle to the fiber bundle
structure under consideration is the statement that the left
invariant metric tensor
$\dot{g}_{ab}$ of any fiber $V_r$ may be transformed by a gauge
transformation $S_0(x)$ to a constant matrix $\dot
{\bar{g}}$ independent of space-time coordinates $x_{\mu}$.

It is easy to see that the gauge symmetry of
$L_{\rm{eff}}\{\widehat{A},\dot{g}_{ab}\}$ leads to the equality
 \begin{equation}\label{eq27}
 L_{\text{eff}}\{\widehat{A}_\mu
,S_0(x)\dot{\bar{g}}_{ab}S_0^{-1}(x)\}=L_{\rm{eff}}\{
\widehat{\overline{A}}_\mu ,\dot{\bar{g}}_{ab}\}
\end {equation}
where  \\
$\widehat{\overline{A}}_\mu =S_0^{-1}(x)\widehat{A}_\mu
S_0(x)-(em)^{-1}S_0^{-1}(x)(\partial _\mu S_0(x))$.

 In other words, in this case the field $\dot{g}_{ab}$ has pure gauge
degrees of
 freedom and there is a gauge condition $\partial_\mu
 (\dot{\bar{g}})_{ab}=0$ when $L_{\rm{eff}}\{\widehat{A},\dot{g}_{ab}\}$
 acquires the simplest form
\begin{multline}\label{eq28}
 L_{\text{eff}}\{\overline{A}_\mu ^a,\dot{\bar{g}}\}=
 -\frac 1{4e_g^2}\widetilde{F}_{\mu
\rho }^a\widetilde{F}_{\nu \sigma }^a\gamma ^{\mu \nu }\gamma
^{\rho \sigma }\\-m_0^2 B_\mu ^s(x)M_{sr}B_\mu ^r(x)+\Lambda
(\dot{\bar{g}}).
\end{multline}
Here
\begin{multline}\label{eq29}
M_{sr}=-\frac 1{v_r}\int\nolimits_{V_r}d\mu (y)\\ \Bigl[
Sp\{(f_s)K(y)\dot{\bar{g}} \overline{K}(y)(f_r)\overline{K}
^{-1}(y)\dot{\bar{g}}^{-1}K^{-1}(y)\}\\-Sp\{(f_s)(f_r)\}\Bigr];\phantom{AA}
\end{multline}
\begin{equation}\label{eq30}
\Lambda (g)=\frac 1{\kappa ^2}\int\nolimits_{V_r}d\mu (y)R^{(f)};
\end{equation}
 and we have introduced new variables
$B_\mu^s(x)=meS_0^{-1}(x)A_\mu^s(x)S_0(x)+S_0^{-1}(x)(\partial_\mu
S_0(x)) $ in order to get the canonical form of the gauge field
strength tensor:
\[
\widetilde{F}_{\mu \nu }^a(x)=\frac 1{e_g}\left( \partial _\mu
B_\nu ^a-\partial _\nu B_\mu ^a+f_{bc}^aB_\mu ^bB_\nu ^c\right).
\]
 In the (\ref{eq28})
\begin{equation*}
 e_g=\left(\frac{Sp(\dot{g})}r \frac{v_r}{m^2\kappa^2}\right)^{-1/2}; \quad
 m_0=\left(\frac{v_r}{\kappa ^2}\right)^{1/2}
\end{equation*}
 We have not considered the gravitation sector of the theory, but it is well
 known \cite{6} that in order to the gravitation sector, which  arises as
a result of deformations of the flat metric $\gamma_{\mu
\nu}$ of the base,
to coincides with ordinary Einstein gravity, one should put
$v_r/\kappa
 ^2=1/16\pi G_N$, where $G_N$ is the Newton gravitational constant.
Therefore it
 is necessary to put $m_0^2=m_p^2/16\pi $, where $m_p$ is the Plank
 mass. The parameter $m^{-1}$, which defines the length scale of
 the extra-dimension manifold, defines only  the value $e_g$ of
 the effective coupling constant
 of the gauge interaction. It is interesting that in the case
 of a simple group $G_r$ the parameter $e$ cancels out.

We are not able to develop here a general theory of the invariant
integration over the Riemannian manifolds with an arbitrary group
of isometries and shell restrict our further consideration to the
case when $V_r$ is a 3-dimensional Riemannian space with the
isometry group of $G_r=SO(3)$.

It is convenient to use the parametrization of $SO(3)$ by the
vectors $y_\alpha $ \cite{7}, where the vector $\vec{y}$
corresponds to the rotation around the axis $\vec{y}/y$ at the
angle $\alpha
$: $y=tan(\alpha /2)$. In this parametrization we have
$f_{ab}^c=\varepsilon _{abc} $,

the Killing vectors of the left shifts are
\begin{equation}\label{eq32}
\xi _{a\alpha }(y)=\frac 12\left( \delta _{a\alpha }+y_ay_\alpha
+\varepsilon _{a\alpha \beta }y_\beta \right)
\end{equation}
and the matrix of the $Ad$-representation is
\begin{equation}\label{eq33}
K_{ab}(y)=\frac 1{1+y^2}\left[ (1-y^2)\delta
_{ab}+2y_ay_b+2\varepsilon _{abc}y_c\right].
\end{equation}
Finally, passing to the spherical coordinates in $V_3$ and taking
into account that $\rm{det}(\xi )=\frac18(1+y^2)^2$, one writes
the left-invariant measure in the form
\begin{eqnarray}
d\mu(y)&=&m^{-3}\sqrt{\dot{g}}\frac 1{\mathrm{{det}(\xi
)}}d^3y\\&=&8m^{-3}\sqrt{
\dot{g}}\frac{y^2}{(1+y^2)^2}dy\sin\Theta d\Theta d\varphi;
\label{eq34} \nonumber\\
v_r&=&\int\nolimits_{V_r}d\mu(y)=8\pi^2m^{-3}\sqrt{\dot{g}}.
\nonumber
\end{eqnarray}
The matrix $\dot{\bar{g}}_{ab}$ is an arbitrary non-degenerate
$(3\times3)$ matrix. It may be diagonalized by a $V$ rotation independent of
$x_\mu $ and has the form
\[
\dot{\bar{g}}_{ab}=V_{aa_1}g_{a_1}\delta
_{a_1b_1}\overline{V}_{b_1b}
\]
In the case of $G_r=SO(3)$ using the invariance of the measure
$d\mu (y)$ and Jacobi  identities for the structure constants
$f_{ab}^c$, it is easy to see that the transformation $B_\mu
^a\rightarrow V_{ab}B_\mu^b$ simultaneously diagonalizes also the
quadratic form
$B_\mu ^aM_{ab}B_\mu ^a$. After the integration of
(\ref{eq29}), we obtain for the eigenvalues of the gauge field
mass matrix:
\begin{multline}\label{eq35}
M_{11}=\frac1{10}\Bigl[ 3\frac{(g_2-g_3)^2}{g_2g_3}\\+\frac{
(g_1-g_3)^2}{g_1g_3}+\frac{(g_1-g_2)^2}{g_1g_2}\Bigr]
\end{multline}
\begin{multline}\label{eq36}
M_{22}=\frac1{10}\Bigl[ 3\frac{(g_1-g_3)^2}{g_1g_3}\\+\frac{
(g_2-g_3)^2}{g_2g_3}+\frac{(g_1-g_2)^2}{g_1g_2}\Bigr]
\end{multline}
\begin{multline}\label{eq37}
M_{33}=\frac1{10}\Bigl[ 3\frac{(g_1-g_2)^2}{g_1g_2}\\+\frac{
(g_3-g_1)^2}{g_3g_1}+\frac{(g_2-g_3)^2}{g_2g_3}\Bigr]
\end{multline}
The integration of the fiber scalar curvature gives
\begin{multline}\label{eq38}
\Lambda (\dot{g}) =-\frac 12\frac{m^2}{\kappa ^2}v_r
\left[2\rm{Sp}(\dot{g}^{-1})-\rm{Sp}(\dot{g}^2)/\rm{det}(\dot{g})\right]
 \\ =\frac{m^2 v_r}{2\kappa ^2}\frac1{g_1g_2g_3}\Bigl[
g_1^2+g_2^2+g_3^2\phantom{AAA}\\-2(g_1g_2+g_1g_3+g_2g_3)\Bigr].
\end{multline}

One can see that for the case of compact $V_3$, (all $g_i$ have
identical signs) the eigenvalues of the gauge field mass matrix
$M_{sr}$ are positive definite. As we see, the values of mass in
the considered approach are determined by the parameters
$(g_i-g_j)$ which characterize the degree of deviation of the
metric from the completely symmetrical case $g_1=g_2=g_3$. As it
is well known, in this case $V_3$ is a symmetric space with the
isometry group $ SO(3)\times SO(3)$, and from
(\ref{eq35}-\ref{eq37}) one can seen that $M_{ii}=0$. If
$g_1=g_2\neq g_3$ then the isometry group of
$V_3$ is $SO(3)\times U(1)$ \cite {8} and in this case
$M_{11}=M_{22}\neq M_{33}$.

 From (\ref{eq38}) it is seen that cosmological constant $\Lambda=0
$ when $g_3=(\sqrt{g_1}\pm\sqrt{g_2})^2$, and this condition does
not impose serious restrictions on $g_i$.

In conclusion, in this paper we have shown how an effective
Lagrangian for the massive gauge fields which does not contain any
additional observable fields can be constructed. Such a Lagrangian
can be rather simply written in the (4 + d)- dimensional form.
(See expr. (\ref{eq21})---(\ref{eq24})). The integration of the
corresponding expressions over extra dimensions in the general
case leads to rather complicated formulae and it is convenient to
perform this integration after imposing appropriate gauge
conditions.

 In this work the mass matrix of the gauge fields has been expressed in
 terms of the left-invariant metric of the additional space $V_r$ and the
calculations
 are made only with the purpose of illustrations of possible results
 for the
 simple case, when the gauge group is $SO(3)$. So we have not
 considered the problem of the determination  of the metric parameters
  and the comparison of obtained expressions with experimental data. We
 did not touch also the problem, why the values $g_i-g_j$ are so small, that
 can reduce the Plank mass $m_p$ down to values of gauge
field masses observable experimentally. It would be interesting
to consider these problems in the framework of ideas of the
spontaneous compactification of multidimensional supergravity
which were intensively studied by the D. V. Volkov group
(\cite{Volkov:1980gb}-\cite{Sorokin:1987zk}).

The author would like to thank I. Bandos and A.A. Zheltukhin for
helpful discussions and interest to this work.

 \end{document}